\documentclass{article}
\usepackage[utf8]{inputenc}
\usepackage{tikz}
\usepackage{qcircuit}
\usepackage{amsmath}
\usepackage{hyperref}
\usepackage{amsthm}
\usepackage{braket}
\usepackage{caption}
\usepackage{subcaption}
\usepackage{graphicx}
\usepackage{lineno}
\usetikzlibrary{arrows.meta,backgrounds}
\usepackage{amssymb}
\usetikzlibrary{positioning}
\usetikzlibrary{arrows,decorations.markings}
\usepackage[backend=biber,style=ieee,sorting=ynt]{biblatex}
\providecommand{\keywords}[1]
{
  \small	
  \textbf{Keywords:} #1
}

\title{On the importance of data encoding in quantum Boltzmann methods}
\author{Merel A. Schalkers\footnote{Corresponding author}  \\
Matthias M\"oller\\
\texttt{\{m.a.schalkers,m.moller\}@tudelft.nl}}
\date{Delft Institute of Applied Mathematics\\[0.5ex]
Delft University of Technology\\[0.7ex]
Mekelweg 4 2628 CD, Delft, The Netherlands}

\begin{document}

\maketitle

\begin{abstract}
In recent years, quantum Boltzmann methods have gained more and more interest as they might provide a viable path towards solving fluid dynamics problems on quantum computers once this emerging compute technology has matured and fault-tolerant many-qubit systems become available.
The major challenge in developing a start-to-end quantum algorithm for the Boltzmann equation consists in encoding relevant data efficiently in quantum bits (qubits) and formulating the streaming, collision and reflection steps as one comprehensive unitary operation. The current literature on quantum Boltzmann methods mostly proposes data encodings and quantum primitives for individual phases of the pipeline assuming that they can be combined to a full algorithm.

In this paper we disprove this assumption by showing that for encodings commonly discussed in literature either the collision or the streaming step cannot be unitary. Building on this landmark result we propose a novel encoding in which the number of qubits used to encode the velocity depends on the number of time steps one wishes to simulate, with the upper bound depending on the total number of grid points.

In light of the non-unitarity result established for existing encodings, our encoding method is to the best of our knowledge the only one currently known that can be used for a start-to-end quantum Boltzmann solver where both the collision and the streaming step are implemented as a unitary operation.
\end{abstract}


\section{Introduction}
Since the first quantum computing boom in the 1990s, quantum computational fluid dynamics (QCFD) has been a field of interest to researchers worldwide. Due to the high computational demands of classic CFD the exponential potential of quantum computers in combination with quantum parallelism and quantum indeterminacy has caused interest in the application. The first QCFD algorithms were proposed by Yepez and his co-workers around the turn of the century \cite{Yepez1998,Yepez2001,YepezBoghosian2001, Yepez2002, Pravia2003}. These algorithms are based on a quantum distributed computing approach assuming that many small-scale quantum computers are more realistic than one large many-qubit system. The core idea of the so-called quantum lattice-gas model is that each grid point of position-space gets its own 6-qubit quantum computer associated to it (which can also be groups of 6 qubits of a future many-qubit quantum computer). The benefit of this approach is that the possible quantum circuit depth and stable entanglement required remains very low, making it a realistic and relatively near-term approach given the capabilities of current quantum devices. Its downside is that to encode a grid of size $N$ a total of $6N$ qubits are required, which means that the amount of qubits required grows linearly with the size of the grid. Given the limited amount of quantum devices available and the large amount of grid points required for solving practical problems with modern Boltzmann methods, this distributed approach proves a significant drawback. Furthermore, as we will show below, the computational basis state encoding of the velocity vector adopted in the aforementioned papers does not allow for implementing the streaming step as a unitary operator so that measurement and state re-initialization is mandatory after each time step.

After these early results by Yepez et al., the QCFD field became stagnant for about a decade until its recent resurgence, in particular, in the form of quantum Boltzmann methods. 
Most recent are the methods presented in \cite{Todorova2020, Budinski2020, Budinski2021, Moawad2022, Schalkers2022, Steijl2023}, that all have their own strengths and weaknesses. Some papers include a streaming and specular reflection mechanism, but no collision methods yet \cite{Todorova2020, Budinski2020, Schalkers2022}. Other approaches have implemented a collision method using the linear combination of unitary approach \cite{Childs2012}, causing the algorithm to require a measurement-and-restart strategy after each time step \cite{Budinski2021}. Due to the high costs of quantum state preparation and the chance of measurement errors this `stop-and-go' strategy is hardly usable in practice. Other algorithms have managed to create a unitary collision operator, but have not yet been able to combine this with a streaming step into one start-to-end algorithm \cite{Moawad2022, Steijl2023}. 

What remained an open problem is the development of a full-fledged quantum Boltzmann method (QBM) that implements both the streaming and the collision step as unitary operations. In this paper we present the first-of-its-kind full-fledged QBM building on a novel encoding scheme of the velocity vector that scales with the number of time steps. Furthermore, we prove rigorously that for the encoding schemes considered for universal quantum computers in all previous publications it is impossible to implement both streaming and collision as a unitary, downgrading them as candidates for any practical QBM. Taking both contributions of this paper together, our new encoding and the theoretical (negative) result on existing encodings, we hope to stimulate a paradigm shift in QBM research from focusing on encodings and algorithms for individual steps of the pipeline to developing full-fledged QBM algorithms. 

\section{Lattice Boltzmann method}
In the Boltzmann method the macroscopic behavior of a fluid is simulated by considering the microscopic behavior of the fluid particles as they move through space and deriving the macroscopic quantities via averaging-based post-processing, instead of encoding the macroscopic variables directly, as is commonly done in other CFD methods like the finite volume method. In recent years the lattice Boltzmann method has gained popularity as it is particularly suited for parallel execution on massively parallel supercomputers. This has led to the development of several open-source packages such as waLBerla \cite{Godenschwager2013,Bauer2021} and openLB \cite{Kummerländer2023} packages. An inclusive overview of the important recent advances in the Lattice Boltzmann method is given in the review paper by Li et al \cite{LinLi2020}. The easy parallelizability of the Boltzmann method also makes it potentially interesting to apply to quantum computers, as the latter allow to efficiently work in a high dimensional space by exploiting quantum parallelism.

\begin{figure}
     \centering
     \begin{subfigure}[b]{0.45\textwidth}
         \centering
         \includegraphics[width=\textwidth]{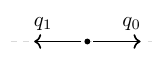}
         \caption{}
         \label{subfig:1a}
     \end{subfigure}
     \hfill
          \begin{subfigure}[b]{0.45\textwidth}
         \centering
         \includegraphics[width=\textwidth]{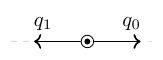}
         \caption{}
         \label{subfig:1b}
     \end{subfigure}
     \hfill
     \begin{subfigure}[b]{0.45\textwidth}
         \centering
         \includegraphics[width=\textwidth]{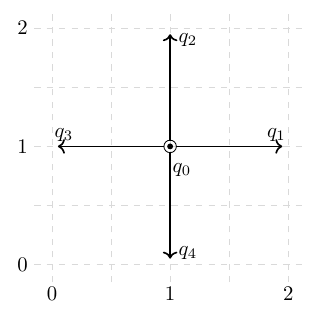}
         \caption{}
         \label{subfig:1c}
     \end{subfigure}
          \hfill
     \begin{subfigure}[b]{0.45\textwidth}
         \centering
         \includegraphics[width=\textwidth]{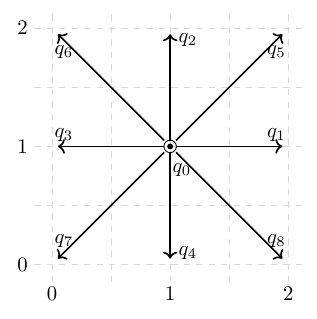}
         \caption{}
         \label{subfig:1d}
     \end{subfigure}
        \caption{Four examples of different types of D$n$Q$m$ possible. Figure \ref{subfig:1a} portrays the D1Q2 setting and Figure \ref{subfig:1b} portrays the D1Q3 setting (where a stationary particle can be included). Figure \ref{subfig:1c} portrays the D2Q5 setting and Figure \ref{subfig:1d} shows the D2Q9 setting. }
        \label{fig:multiple_DnQm}
\end{figure}

In this paper we consider the discrete lattice Boltzmann method, where a particle can only move with specific velocities taken from a finite set of discrete velocities. We define the structure of the method using the D$n$Q$m$ system, where $n$ represents the amount of spatial dimensions and $m$ the amount of discrete velocities considered. Figure \ref{fig:multiple_DnQm} gives examples of the commonly used D1Q2, D1Q3, D2Q5 and D2Q9 systems, respectively, in standard Boltzmann convention. For an in depth review of the lattice Boltzmann method we refer to the book \cite{Kruger2017}.

Boltzmann methods simulate the macroscopic behavior of a fluid or gas by implementing a streaming step followed by particle collision on the microscopic level in each time step. When obstacles are present an additional reflection step is performed in each time step. For brevity we omit a detailed description of the latter and refer the interested reader to our recent work \cite{Schalkers2022} on this topic.

\begin{figure}
     \centering
     \begin{subfigure}[b]{0.5\textwidth}
         \centering
         \includegraphics[]{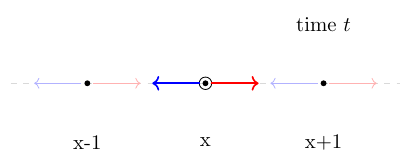}
         \caption{}
         \label{subfig:2a}
     \end{subfigure}
     \hfill
     \begin{subfigure}[b]{0.5\textwidth}
         \centering
         \includegraphics[]{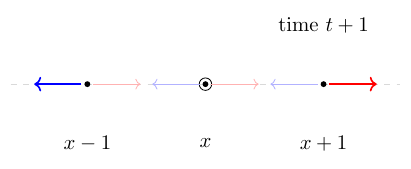}
         \caption{}
         \label{subfig:2b}
     \end{subfigure}
        \caption{Illustration of the streaming step for the D1Q3 case. Figure \ref{subfig:2a} shows the velocity vectors at position $x$ at time $t$. Figure \ref{subfig:2b} shows the same after configuration at time $t+1$ after particles have moved to positions $x-1$ and $x+1$, respectively. Red and blue colors identify the different streaming directions and their propagation pattern.}
        \label{fig:D1Q2_streaming}
\end{figure}

The streaming step is implemented by letting the particles move by one grid point per time step in the direction they are traveling currently. Figure \ref{fig:D1Q2_streaming} illustrates how the particles travel in one time step from the point $x$ to $x\pm 1$ respectively for the D1Q3 case. Similar illustrations can be constructed for two- and three-dimensional cases but are omitted here for brevity reasons.

\begin{figure}
     \centering
     \begin{subfigure}[b]{0.45\textwidth}
     \caption{}
     \label{subfig:3a}
         \centering
\includegraphics[]{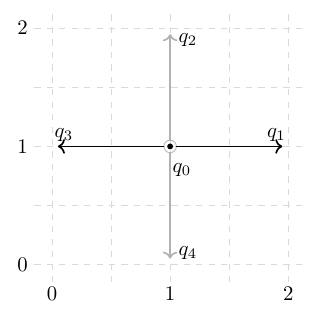}
     \end{subfigure}
     \hfill
     \begin{subfigure}[b]{0.45\textwidth}
     \caption{}
     \label{subfig:3b}
         \centering
\includegraphics[]{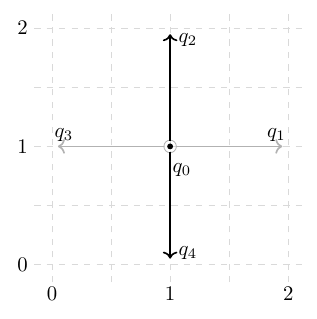}
     \end{subfigure}
\caption{Illustration of two velocity combinations of the D2Q5 (and D2Q4) velocity spectrum that belong to the same equivalence class with total momentum 0 and mass 2: \ref{subfig:3a} particles streaming in the $q_1$ and $q_3$ direction, and \ref{subfig:3b} particles streaming in the $q_2$ and $q_4$ direction.}
\label{fig:equivalence_class}
\end{figure}

To implement the collision step we define so-called equivalence classes of streaming patterns which have the same total mass and momentum and are thus considered to be equivalent. A combination of colliding particles can therefore be transformed into any combination from the same equivalence class upon collision without changing the total mass and momentum. Figure \ref{fig:equivalence_class} shows an example of two equivalent velocity combinations for the D2Q5 (and D2Q4) case.

In Section \ref{sec:proof}, we provide rigorous mathematical proofs that show that such a unitary treatment of both streaming \emph{and} collision is impossible with the encodings adopted in current literature, thereby underpinning the uniqueness and urgent need of our proposed space-time encoding.
Subsequently, in Section \ref{sec:novel_encoding}, we present a lattice Boltzmann encoding for which both the collision and the streaming step can be performed through unitary operations and thus admit a straightforward implementation on a sufficiently large fault-tolerant quantum computer.

\section{Data encoding}\label{sec:proof}
As in any computational field, data encoding is pivotal for reaching a good result. More than five decades of classical CFD research and application have established `good practices' for storing field data such as densities and velocities at, e.g., the grid points or cell centers as floating-point numbers following the IEEE-754 standard. Every now and then new hardware developments stimulate research into non-standard formats, like reduced or mixed-precision \cite{Freytag2022}, but, in general, data encoding is not considered to be an open problem. 

Not so in QCFD and, in particular, quantum Boltzmann methods where different encoding methods are used in different papers. The two mainstream encodings of the velocity vector are the amplitude based encoding \cite{Todorova2020, Budinski2020, Budinski2021, Schalkers2022} and the computational basis state encoding \cite{Yepez1998, Yepez2001, YepezBoghosian2001, Yepez2002, Pravia2003, Moawad2022, Steijl2023}. In this section we will review the main data encodings currently used for QBM and show that in all of them either the streaming step or the collision step cannot be unitary. This result, though discouraging at first sight, should be interpreted as wake-up call that novel quantum encodings for CFD states are imperative for devising full-fledged QCFD applications in the future. We propose one such novel encoding in Section \ref{sec:novel_encoding} and discuss its potential and limitations.

The two mainstream encodings of the velocity vector are the amplitude based encoding \cite{Todorova2020, Budinski2020, Budinski2021, Schalkers2022} and the computational basis state encoding \cite{Yepez1998, Yepez2001, YepezBoghosian2001, Yepez2002, Pravia2003, Moawad2022, Steijl2023}. In what follows, we will consider both approaches separately and show how they both lead to a contradiction in the unitarity of either the collision or the streaming operation. 

\subsection{Amplitude based encoding}\label{ssec:amplitude_based_encoding}
The first type of encoding we consider is the so-called amplitude based encoding, used for several quantum Boltzmann methods \cite{Todorova2020, Budinski2020, Budinski2021, Schalkers2022}.
The amplitude based encoding of the velocity vector is such that at each location $\ket{x}$ there can be multiple particles with different velocities, for instance $\ket{v_0}$, $\ket{v_1}$, $\ket{v_2}$ and $\ket{v_3}$ for D2Q4. Here and below, $\ket{i}$ denotes the representation of $i$ as bit string. The state of the system at this point $x$ can then be encoded as\footnote{Note that we distinguish in our notation between the grid point $x$ and its representation $\ket{x}$ as part of the quantum register.}
\begin{equation}
\ket{x}\left (\alpha_0\ket{v_0} + \alpha_1\ket{v_1} + \alpha_2\ket{v_2} + \alpha_3\ket{v_3} \right ),
\end{equation}
where $\alpha_0$, $\alpha_1$, $\alpha_2$ and $\alpha_3$ are complex numbers that simply represent the relative weight or amount of particles traveling at the given velocity at grid point $x$. For simplicity we will assume that $|\alpha_0|^2+|\alpha_1|^2+|\alpha_2|^2+|\alpha_3|^2=1$, and so in this example there are only particles at grid point $x$ but the proof extends trivially to the general case with particles spread around the grid.

In order to show that this encoding of the velocity vector inevitably leads to non-unitary collision operators, let us first take a close look at what is required from a collision operator. A collision operator $U_\text{col}$ needs to map the velocities of the incoming particles to a so-called equivalent outgoing state. Two states are equivalent if the total mass and momentum of all particles combined are the same. For example having one particle of mass 1 traveling in the positive $x$ direction and one in the negative $x$ direction is equivalent to 1 particle of mass 1 traveling in the positive $y$ direction and 1 particle of mass 1 in the negative $y$ direction, as for both states the total momentum is 0 and the mass is the same. This means that if there is only one particle with one direction at a specific point in space, the collision operator cannot change this as there is no other direction this one particle could be traveling at that has the same total momentum. Using these requirements we can set up a generic collision operation $U_\text{col}$ that solely meets the basic requirements of behavior it needs to portray. The first requirement is that there should be at least a combination of incoming velocity states that leads to a new combination containing at least one velocity state that was previously not present. Let the state $\ket{\psi_1}$ be an example of an incoming state for which a state in its equivalence class includes at least some velocity which is not included in the original state. Without loss of generality assume that $\ket{\psi_1}$ consists of two different velocity states $\ket{v_0}$ and $\ket{v_1}$, meaning that $|\alpha_0|,|\alpha_1| > 0$ and $\alpha_2=\alpha_3=0$. Then we can write the state of the system as
\begin{equation}\label{eq:sys1}
\ket{\psi_1} = \ket{x}\left (\alpha_0\ket{v_0} + \alpha_1\ket{v_1} \right ).
\end{equation}
Now assume that an equivalent velocity combination exists consisting of particles traveling with velocities $\beta_2\ket{v_2} + \beta_3\ket{\psi_3}$, where we have $|\beta_2|,|\beta_3| > 0$ and we let $\ket{\psi_3}$ be any combination of all basis states except $\ket{v_2}$. To realize this potential outcome of a collision as a quantum algorithm, we need to implement the transformation between both equivalent states as a unitary operation $U_\text{col}$ which changes the states of the velocity encodings as follows
\begin{equation}
\begin{split}
     \ket{\psi_1^\prime} &= I \otimes U_\text{col} \ket{\psi_1} \\& = \ket{x}\otimes U_\text{col}\left (\alpha_0\ket{v_0} + \alpha_1\ket{v_1} \right ) \\ & =\ket{x} \left ( \gamma_0(\alpha_0\ket{v_0} + \alpha_1\ket{v_1}) + \gamma_1(\beta_2\ket{v_2} + \beta_3\ket{\psi_3})  \right ).
\end{split}
\end{equation} 
Here, if $\gamma_0=1$ and $\gamma_1=0$ no collision is taking place (and we simply implement an identity operation) and if $\gamma_1=1$ we fully change from the original velocities to its alternative representative from the same equivalence class.\footnote{Here $\gamma_0$, $\gamma_1$ are chosen to reflect the fact that a collision operation should switch weight of a combination of velocities in an equivalent class to another combination of velocities in the same equivalence class. This equation could be written in a less restrictive way by splitting $\gamma_0$ and $\gamma_1$ up into separate amplitudes $\gamma_i$ for all the basis states $\ket{v_i}$, the same contradiction of unitarity as presented below however could be reached.} Note that to preserve unitarity $|\gamma_0|^2+|\gamma_1|^2=1$ must hold.

Let us now consider another system in state $\ket{\psi_2} = \ket{x}\ket{v_2}$. Applying the unitary operation $U_\text{col}$ should not effect the state at all as a single speed is only in an equivalence class with itself, and so the required behavior for $U_\text{col}$ is
\begin{equation}\label{eq:sys2_end}
\begin{split}
     \ket{\psi_2^\prime} &=I \otimes U_\text{col}\ket{\psi_2} \\ & =  \ket{x}U_\text{col}\ket{v_2} \\ & = e^{i\theta}\ket{x}\ket{v_2},
\end{split}
\end{equation}
with $\theta\in (0,2\pi]$. That is, the collision operator must preserve the single-velocity state except for changes in the phase factor $e^{i\theta}$ that can be neglected.

Now that we have identified the required behavior for $U_\text{col}$ to implement a collision operation, we can prove that any $U_\text{col}$ that meets both requirements simultaneously cannot be unitary. Here, we resort to the characterization $U_\text{col}^\dagger U_\text{col}=I$ of unitary operators, with superscript $\dagger$ denoting the adjoint operator.

\begin{proof}
To reach a contradiction, assume that $U_\text{col}$ is a unitary operator. Then it must preserve the inner product for all possible states $\ket{\phi_1}$ and $\ket{\phi_2}$
\begin{equation}
    \braket{\phi_1 | \phi_2} = \bra{\phi_1} U_\text{col}^\dagger U_\text{col} \ket{\phi_2}.
\end{equation}

However, for a collision operation $U_\text{col}$ that behaves as expected on the system states described in Equations \eqref{eq:sys1} to \eqref{eq:sys2_end}, it follows that 
\begin{equation}
\begin{split}
0 &=\braket{\psi_1|\psi_2} \\ & = \bra{\psi_1} \left ( I \otimes U_\text{col} \right )^\dagger \left ( I \otimes U_\text{col} \right ) \ket{\psi_2} \\ & = e^{i\theta}\left ( \gamma_0(\alpha_0\bra{v_0} + \alpha_1\bra{v_1}) + \gamma_1(\beta_2\bra{v_2} + \beta_3\bra{\psi_3})  \right )  \bra{x} \ket{x}  \ket{v_2}  \\ & = e^{i\theta}\gamma_1\beta_2.
\end{split}
\end{equation}
The first equality follows from the fact that $\ket{\psi_1}$ and $\ket{\psi_2}$ are orthogonal by construction. The second one holds under the assumption of $U_\text{col}$ being unitary, which is disproved by the fact that the entire equality chain only holds for the trivial case $\gamma_1=0$ (as $|\beta_2|>0$ by definition of the state $\ket{\psi_1}$), that is, when $U_\text{col}$ does not implement the collision operation.
From this we can conclude that an amplitude based encoding of the velocity does not allow for a unitary implementation of the collision operation. 
\end{proof}
Notice that this proof works for any amplitude based encoding of $v$ where the different possible velocities at a position are all represented by their own basis state as there will always be a case with only a single incoming velocity, for which an identity operation up to a phase shift should take place, while at the same time there will be combinations of velocities for which we want some weight of the system to change from one combination of velocities to another combination of velocities in the same equivalence class. These two antagonizing requirements will always lead to the same contradiction of unitarity proven above and we further expand on this intuition in Section \ref{ssec:intuition}.

\subsection{Computational basis state encoding}\label{ssec:comp_basis_encoding}
The second type of encoding of a quantum state considered is the computational basis encoding, used in several quantum lattice Boltzmann papers such as \cite{Yepez1998, Yepez2001, YepezBoghosian2001, Yepez2002, Pravia2003, Moawad2022, Steijl2023}. Using this encoding the contradiction of unitarity in the collision operation can be avoided by encoding the velocity of the qubits at a position $\ket{x}$ in space by identifying each direction particles could be streamed from with its own qubit, which will be set to one if and only if there is a particle streaming from that direction.

\begin{figure}[h]
\centering
    \includegraphics[]{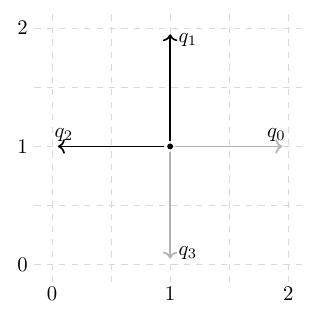}
    \caption{Illustration of the computational basis state encoding for the D2Q4 lattice. For each grid point $x$ we set the respective qubit $q_j$ to one if and only if there is a particle streaming in that direction, i.e. $\ket{v} = \ket{q_0q_1q_2q_3} = \ket{0110}$.}
    \label{fig:D2Q4_ex}
\end{figure}

As an example consider the D2Q4 lattice depicted in Figure \ref{fig:D2Q4_ex}. In this case the velocity can be encoded using four qubits $q_0$, $q_1$, $q_2$ and $q_3$ where the state
\begin{equation}
    \ket{x}\ket{v} = \ket{x}\ket{q_0q_1q_2q_3} = \ket{x}\ket{0110}
\end{equation}
is such that from the center point $(1,1)$, there is a particle streaming to $(1,2)$ and a particle streaming to $(0,1)$ but not to $(2,1)$ or $(1,0)$. 

Using this encoding the collision step can be defined quite naturally as unitary operation. However, we run into trouble when attempting to define a unitary streaming step $U_\text{str}$ as we demonstrate in what follows. 

\begin{figure}
    \centering
     \begin{subfigure}[b]{\textwidth}
         \centering
    \includegraphics[]{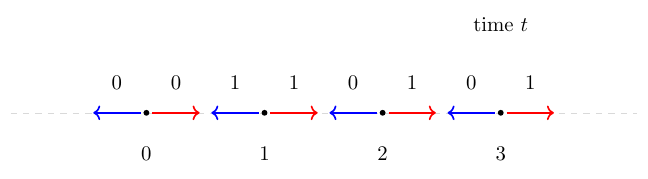}
         \label{fig:D1Q2ex_1_1}
    \vspace{.5cm}
     \end{subfigure}
          \begin{subfigure}[b]{\textwidth}
          \centering
\includegraphics[]{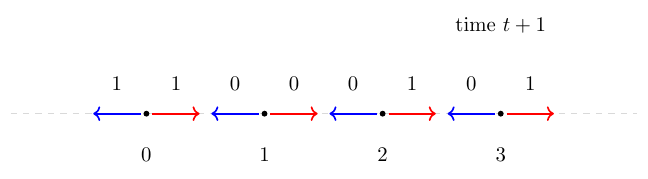}
         \label{fig:D1Q2ex_1_2}
     \end{subfigure}
    
    \caption{D1Q2 example setting 1. The binary encoding above the arrows indicate whether or not a particle is flowing there in that time step. 1 indicates that there is a particle streaming there and 0 indicates that there is no particle. In the example setting we consider periodic boundary conditions. The top figure shows the state of the system at time $t$. The figure below shows the state of the system at time $t+1$.}
    \label{fig:D1Q2ex_1}
\end{figure}

\begin{figure}
         \centering
     \begin{subfigure}[b]{\textwidth}
         \centering
        \includegraphics[]{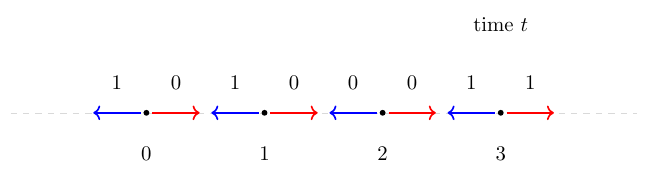}
         \label{fig:y equals x}
    \vspace{.5cm}
     \end{subfigure}
     \begin{subfigure}[b]{\textwidth}
         \centering
\includegraphics[]{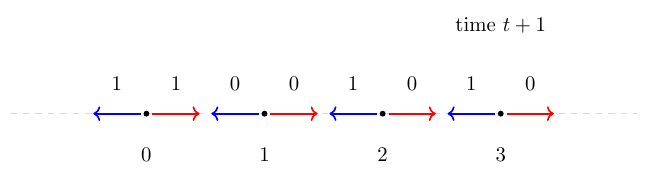}
         \label{fig:three sin x}
     \end{subfigure}
    \caption{D1Q2 example setting 2, the binary encoding above the arrows indicate whether or not a particle is flowing there in that time step. 1 indicates that there is a particle streaming there and 0 indicates that there is no particle. In the example setting we consider periodic boundary conditions. The top figure shows the state of the system at time $t$. The figure below shows the state of the system at time $t+1$.}
    \label{fig:D1Q2ex_2}
\end{figure}

To simplify notation let us restrict ourselves to the D1Q2 lattice and consider the two settings at time $t$ from Figures \ref{fig:D1Q2ex_1} and \ref{fig:D1Q2ex_2}, which can be encoded as
\begin{align}
    \ket{\psi_1} &= \sum_{x=0}^3 \ket{x}\ket{v}\\ &= \frac{1}{2}\left ( \ket{00}\ket{00} + \ket{01}\ket{11} + 
\ket{10}\ket{10} + \ket{11}\ket{10}\right ),
\label{eq:psi1_str}
\end{align}
and 
\begin{equation}
    \ket{\psi_2} = \frac{1}{2} \left ( \ket{00}\ket{01} + \ket{01}\ket{01} + 
\ket{10}\ket{00} + \ket{11}\ket{11} \right ),
\label{eq:psi2_str}
\end{equation}
respectively. It then follows directly that 
\begin{equation}
    \braket{\psi_1|\psi_2} = 0.
    \label{eq:orthogonal}
\end{equation}

Upon streaming, the systems from Figures \ref{fig:D1Q2ex_1} and \ref{fig:D1Q2ex_2} change from their state at time $t$ (top lattice) to that at time $t+1$ (bottom lattice), i.e.
\begin{equation}
    \ket{\psi_1^\prime} = \frac{1}{2} \left ( \ket{00}\ket{11} + \ket{01}\ket{00} + 
\ket{10}\ket{10} + \ket{11}\ket{10} \right ),
\label{eq:psiprime1_str}
\end{equation}
and 
\begin{equation}
    \ket{\psi_2^\prime} = \frac{1}{2} \left ( \ket{00}\ket{11} + \ket{01}\ket{00} + 
\ket{10}\ket{01} + \ket{11}\ket{01} \right ) ,
\label{eq:psiprime2_str}
\end{equation}
respectively. As in the previous section, we will show by contradiction that any operation $U_\text{str}$ for which $U_\text{str}\ket{\psi_1} = \ket{\psi_1^\prime}$ and $U_\text{str}\ket{\psi_2} = \ket{\psi_2^\prime}$ cannot be unitary. 

\begin{proof}
    Let us assume that $U_\text{str}$ is unitary, i.e. it preserves the inner product
    \begin{equation}
        \braket{\phi_1 | \phi_2} = \bra{\phi_1} U^\dagger_\text{str} U_\text{str} \ket{\phi_2},
    \end{equation}
for all states $\ket{\phi_1}, \ket{\phi_2}$. Substituting the states \eqref{eq:psi1_str} and \eqref{eq:psi2_str} on the left side, and \eqref{eq:psiprime1_str} and \eqref{eq:psiprime2_str} into the right inner product we arrive at the contradiction 
\begin{equation}
    0=\braket{\psi_1 | \psi_2} = \bra{\psi_1}U_\text{str}^\dagger U_\text{str} \ket{\psi_2} =
    \braket{\psi_1^\prime | \psi_2^\prime} = \frac{1}{2}.
\end{equation}
The first equality follows from the orthogonality property \eqref{eq:orthogonal}, and the second one from the assumption that $U_\text{str}$ is a unitary operator, which we just disproved.
\end{proof}

As in Section \ref{ssec:amplitude_based_encoding} this proof extends to any computational basis encoding where each possible combination of velocities at a specific lattice point is encoded using its own basis state, as one can always construct two situations with no overlap at time $t$ that will have non-zero overlap after streaming at time $t+1$. This proof also extends trivially to any other D$n$Q$m$ setting as the streaming possibilities of D1Q2 are essentially a subset of any other system and thus the same example can be used by setting the other streaming directions to 0.

\subsection{Intuition and extension of non-unitarity proofs}\label{ssec:intuition}
In this section we expand on our non-unitarity proofs by providing physical intuition behind the proofs presented above. It is intended to give insight into what types of encodings our non-unitarity proof extends to, and what physical features of the system necessarily lead to the non-unitarity for these encodings. 

Consider the proof from Section \ref{ssec:amplitude_based_encoding} that shows that the amplitude based encoding, where each velocity direction is identified through its own basis state leaving the total velocity at a position $x$ to be a superposition of such basis states, prevents the collision operator $U_\text{col}$ from being unitary. Since it encodes each streaming direction as a different basis state, the quantum encodings of the velocity directions are all orthogonal to one another. This is also necessary, since if the basis states of the possible streaming directions are not orthogonal, we cannot fully distinguish between them. However, this orthogonality of the different velocity directions leads directly to the non-unitarity of $U_\text{col}$. Since a collision operator that will rotate a given linear combination of basis states into a linear combination of other basis states in such a way that the represented streaming patterns belong to the same equivalence class, it will also rotate `pure' velocities represented by a single basis state into another basis state leading to a nonphysical and undesired change of velocities. 

Following this line of argumentation it can be seen that the non-unitarity of $U_\text{col}$ is not so much a result of a specific choice of encoding but an inherent non-unitarity of the collision step itself that directly leads to the idea of computational basis state encoding, where each velocity pattern (i.e. the combination of velocities) at a grid point is encoded as its own basis state, and not as a unitary combination of all the basis states representing a non-zero contribution.

When encoding the velocity pattern at each grid point as a basis state, naturally, the non-unitarity of collision falls away and we can find a straightforward unitary operator to implement the collision step.
However, such an encoding will always lead to non-unitarity of streaming due to the non-local nature of a streaming operation. Consider an arbitrary point in space $x$ and imagine two different scenarios with two different combinations of speeds $\ket{v_1}$ and $\ket{v_2}$ at this point. Then the inner product between $\ket{x}\ket{v_1}$ and $\ket{x}\ket{v_2}$ must be 0, as these are different basis states. However, the velocity states of the systems at position $x$ in the next time step do not depend on the current velocity states in the lattice point. In fact, they only depend on the velocity states of the neighboring lattice points. Since the inner product of the states at the point $x$ at the next time step does not depend on the current states at the point $x$, in the next time step the velocity at the point $x$ of the two systems could be identical, and hence, the inner product could be one. There is no way of ensuring that this can only happen when the inner product at some other point $x^\prime$ of the systems was non-zero before as each grid point has velocity vectors in multiple directions determining its associated velocity basis state. 

This shows that any quantum encoding that successfully implements both streaming and collision as a unitary operation must belong to one of the following three types. The first type is an amplitude based type encoding, where the different velocities are not orthogonal and thus not entirely distinguishable. The second type is a computational basis state encoding where the non-locality of streaming is somehow avoided. The last type is a completely novel encoding method that avoids both non-unitarity problems entirely.
In the next section we will present precisely one such idea.

\section{Space-time data encoding}\label{sec:novel_encoding}
In this section we propose a novel space-time data encoding that enables unitary collision \emph{and} streaming at the same time. To the best of our knowledge, this is the first-of-its-kind start-to-end quantum Boltzmann algorithm that does not require measurement and quantum-state re-initialization after each time step.

In what follows, we adopt an extended computational basis state encoding, where at each location $x$ we take into account the velocities at all grid points in the vicinity of $x$. Here, `in the vicinity of $x$' means that a particle can theoretically reach the grid point $x$ within the number of time steps still to be performed before measurement. Mathematically speaking being `in the vicinity of $x$' means being, respectively, in the so-called extended von Neumann, Moore or hexagonal neighborhood of the point $x$, depending on the lattice structure.\footnote{The von Neumann neighborhood of extent $r$ defines the diamond-shaped set of points at a Manhattan distance of up to $r$ from the point $x$. It applies to, e.g., D2Q4, D2Q5, D3Q6, and D3Q7. The Moore distance extends the former one by diagonal directions and applies to, e.g., D2Q8, D2Q9, D3Q26, and D3Q27. As its name suggests, the hexagonal neighborhood applies to D2Q6 and its extension to its three-dimensional counterpart.} 

This leads to a trade-off between the number of time steps that can be performed between measurements and the number of qubits required to encode the velocity at each grid point $x$. The more time steps one wishes to take between measurement-and-re-initialization cycles, the more qubits are required for our space-time encoding. Obviously the maximum number of qubits required to implement the velocity without any in-between measurements must be such that the entire grid is spanned. For a D$n$Q$m$ lattice this will be $mN_g$, where $N_g$ is the total number of grid points. When encoding the proposed method on a classical computer $mN_g$ bits would also be required, so when encoding the full domain there is no quantum benefit in terms of (qu)bit numbers. The quantum improvement comes from exploiting quantum parallelism, which is done as long as we do not encode the whole space.

In what follows, let $N_t$ denote the number of streaming steps to be performed between (re-)initialization and measurement. We extend the computational basis state encoding of velocity directions from Section \ref{ssec:comp_basis_encoding} to take into account all the speed states from grid points in the neighborhood of $x$ that can (at least theoretically) reach $x$ within $N_t$ streaming steps. This takes away the non-locality of the streaming operator, which led to the non-unitarity of $U_\text{str}$ for the `regular' computational basis state encoding at the cost of increasing the number of qubits required to encode all required velocity data.

\begin{figure}[h]
\centering
    \includegraphics[]{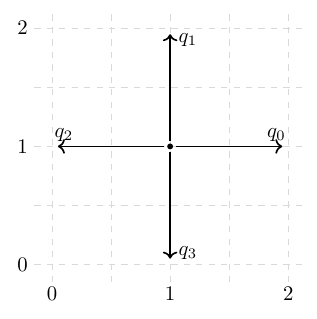}
    \caption{Illustration of the computational basis state encoding for D2Q4.}
    \label{fig:D2Q4}
\end{figure}

\begin{figure}[h]
\centering
    \includegraphics[]{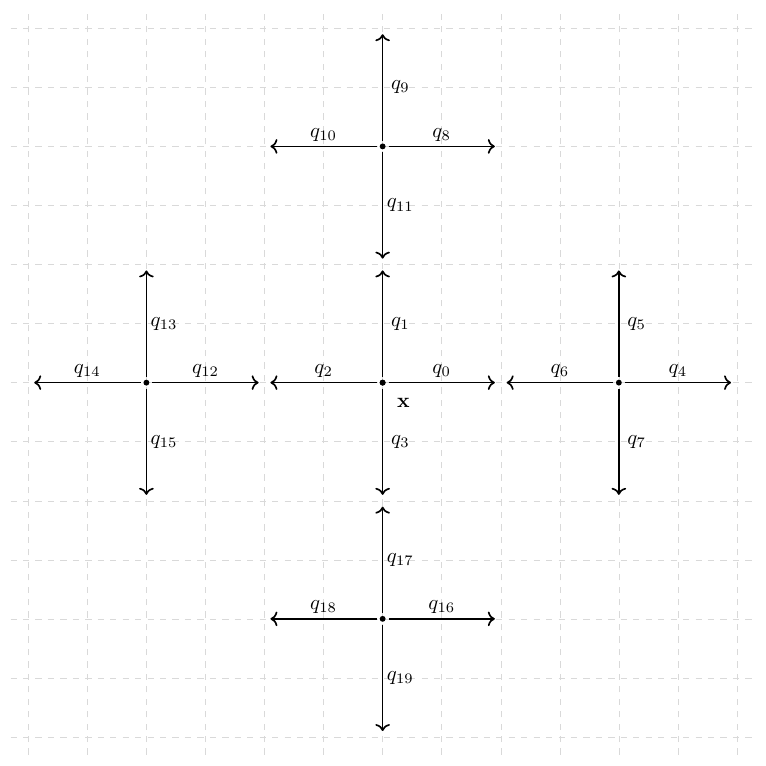}
    \caption{Illustration of the space-time encoding for D2Q4 for a single time step.}
    \label{fig:D2Q4_quantum}
\end{figure}

We will give a detailed description of this encoding for the D2Q4 lattice, but want to note that it can be extended naturally to any other choice of D$n$Q$m$. Consider the D2Q4 lattice given in Figure \ref{fig:D2Q4} with qubit $q_j$ set to one if and only if there is a particle traveling with velocity direction $j$ from grid point $x$ into a neighboring grid point in the current time step. We now extend this encoding to include \emph{all} possible velocities at positions `in the vicinity of $x$' for the total of $N_t$ time steps in order to obtain a unitarily streamable encoding. This is illustrated in Figure \ref{fig:D2Q4_quantum} for a single time step, i.e. $N_t=1$ yielding the encoding
\begin{equation}
    \ket{x}\ket{q_{19}q_{18} \dots q_0}.
\end{equation}
For D2Q4, the number of qubits encoding the possible velocity states per grid location $x$ grows with the number of time steps (still) to be taken as
\begin{equation}\label{eq:von_Neumann}
    n_v = 4 + \sum_{i=1}^{N_t}16i = 8N_t^2 + 8N_t + 4,
\end{equation}
where the maximum number of qubits required to encode all velocity directions over the entire grid equals $4N_g$ as stated before.\footnote{Note that the growth rate of qubit numbers per time step depends on the choice of D$n$Q$m$. The number of qubits required is equal to the number of points in the extended Von Neumann, Moore or hexagonal neighborhood, depending on which choice of $n$ and $m$ considered.}
Similarly it can be shown that for $d$ dimensions the growth rate is of the order $\mathcal{O}\left (N_t^d \right)$.

\begin{figure}
    \centering\includegraphics{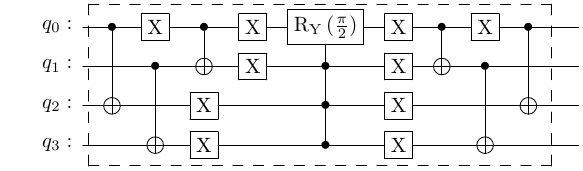}
    \caption{An example of an implementation of the collision operation $U_{col}$ for the D2Q4 example with $|\alpha|=|\beta|=\frac{1}{\sqrt{2}}$.}
    \label{fig:U_col}
\end{figure}

We can now encode the collision step by first identifying the equivalence class for the D2Q4 lattice. We note that at each grid point $x$ as represented in Figure \ref{fig:D2Q4} the states $\ket{q_0q_1q_2q_3} = \ket{1010}$ and $\ket{q_0q_1q_2q_3} = \ket{0101}$ belong to the same equivalence class (cf. Figure \ref{fig:equivalence_class}), as they have the same total mass and momentum.\footnote{The other equivalence classes are $\ket{q_0q_1q_2q_3} = \ket{1000}$ and $\ket{q_0q_1q_2q_3} = \ket{1100}$ and all cyclic shifts of these patterns, and $\ket{q_0q_1q_2q_3} = \ket{1111}$. However, they all have just a single representative so that we define the collision operator based on the ambiguous case.} We implement the collision step by defining a unitary operator $U_\text{col}$ which performs the following mappings 
\begin{align}
    U_\text{col}\ket{1010} &= \phantom{-}\alpha\ket{1010} + \beta\ket{0101},\\
    U_\text{col}\ket{0101} &= -\beta\ket{1010} + \alpha\ket{0101},
\end{align} 
with $\alpha, \beta \in \mathbb{C}$ and $|\alpha|^2+|\beta|^2=1$,
while acting as the identity operation on any other basis state.\footnote{This operation is unitary, as can be verified by writing it as 
\begin{equation}
U_\text{col} = \Pi_{0000,1010} \Pi_{0000,0101} \begin{bmatrix}
    \alpha & \beta & 0 & \dots & 0 \\
    -\beta & \alpha & 0 & \dots & 0 \\
    0 & 0 & 1 & \dots & 0 \\
    \dots & \dots & \dots & \dots & \dots \\
    0 & 0 & 0 & 0 & 1
\end{bmatrix} \Pi_{0000,0101}\Pi_{0000,1010}.\end{equation} Here the matrices $\Pi_{i,j}$ represent permutation matrices between the basis states $i$ and $j$ which are trivially unitary. Since the product of unitary matrices is a unitary matrix we only need to show the unitarity of the matrix 
\begin{equation}
    M = \begin{bmatrix}
    \alpha & \beta & 0 & \dots & 0 \\
    -\beta & \alpha & 0 & \dots & 0 \\
    0 & 0 & 1 & \dots & 0 \\
    \dots & \dots & \dots & \dots & \dots \\
    0 & 0 & 0 & 0 & 1
\end{bmatrix},
\end{equation} to show the unitarity of $U_\text{col}$. 
The unitarity of M follows directly from writing out $M^\dagger M $ and finding $I$. } With the so-defined $U_\text{col}\in \mathbb{C}^{2^4 \otimes 2^4}$, we can write the total collision operation for an encoding of the velocities states $v$ consisting of $n_v=4k$ qubits as $k$-fold Kronecker products of $U_\text{col}$ operations, i.e.  $U_\text{col}^\text{tot}=U_\text{col} \otimes \dots \otimes U_\text{col}$. Since each $U_\text{col}$ requires a few CNOT and a single triple controlled rotation gate, see Figure \ref{fig:U_col}, the total collision operator can be efficiently implemented even on near-term devices.\footnote{We can implement the described collision operator by first applying three CNOT operations to the system turning the states into $\ket{1010} \mapsto \ket{1110}$ and $\ket{0101} \mapsto \ket{1111}$. Subsequently a triple controlled rotation operation of choice is applied to the right-most qubit (controlled on the three left-most qubits). Finally the initial three CNOT operations are applied in reverse order to reset all velocity states correctly.}

In practice the total collision operator $U_\text{col}^\text{tot}$ differs per time step, since its local counterpart $U_\text{col}$ only needs to be applied to velocity states `in the vicinity of $x$'. 
In the first out of the $N_t$ time steps it is important for all qubits representing velocity states `in the vicinity of $x$' to be updated correctly. In the very last time step, however, it is only important for the qubits $q_0$, $q_1$, $q_2$ and $q_3$ to end up in the correct state. The more time steps $t$ have been taken, the less time steps $N_t-t$ are still to be taken and so the 4-qubit local collision operator $U_\text{col}$ only needs to be applied to the remaining qubits relevant for encoding the `directly connected' velocity states as given in Equation \eqref{eq:von_Neumann}. 

\begin{figure}
    \centering
    \includegraphics{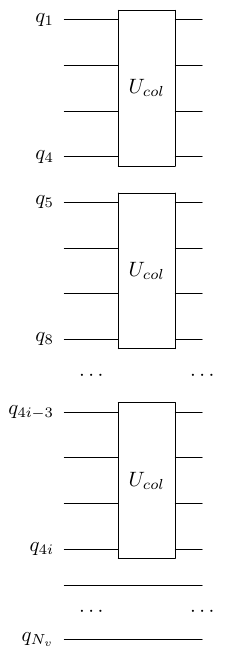}
    \caption{The collision operation applied in the i-th time step for the D2Q4 example.}
    \label{fig:collision_i_timesteps}
\end{figure}

With this logic we can define a collision operator per time step $t$ as
\begin{equation}
U_{\text{col},t}^\text{tot}=\underbrace{U_\text{col} \otimes \dots \otimes U_\text{col}}_{\text{$c$  collision operations}} \otimes \underbrace{I \otimes \dots \otimes I}_{\text{identity operations}},
\end{equation}
where $c = 2(N_t-t)^2 + 2(N_t-t) +1$ and the identity operations are added to avoid dimensionality issues. In practice no operation will be applied on the qubits encoding velocity states not `in the vicinity of $x$' within $N_t-t$ time steps, Figure \ref{fig:collision_i_timesteps} shows what this looks like as a quantum circuit.

\begin{figure}
    \centering
    \includegraphics{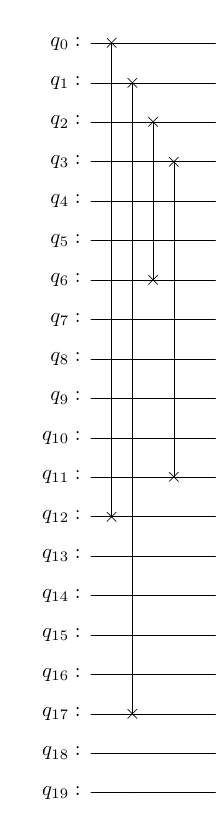}
    \caption{An example of an implementation of streaming in the D2Q4 case with $t=1$.}
    \label{fig:streaming_impl}
\end{figure}

Our space-time encoding enables different manners of implementing the streaming step. It can easily be seen that the way the streaming method should be implemented differs per time step $t$ depending on which positions will be `in the vicinity of $x$' in the next time step as well. At the first time step it is important for (almost) all qubits to be streamed to a very specific position, whereas in the last time step it is only important for the qubits $q_0$, $q_1$, $q_2$ and $q_3$ to end up in the correct state. For the example shown below we are only considering a total of one step to be taken (i.e. $N_t=1$) and so we only need to consider the speeds that will stream to location $x$ in one time step. In this case that means that streaming consists of performing a swap operation between the following qubit pairs $q_0$ and $q_{12}$, $q_1$ and $q_{17}$, $q_2$ and $q_6$ as well as $q_3$ and $q_{11}$, see Figure \ref{fig:streaming_impl}.  

\begin{figure}
    \centering
    \includegraphics{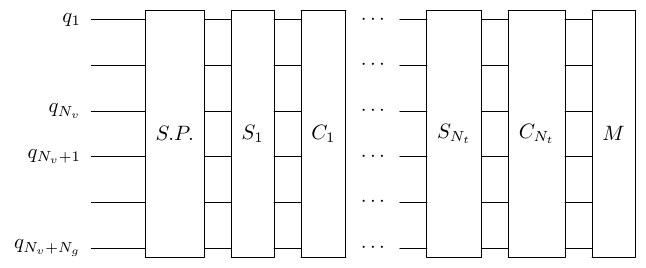}
    \caption{The full space-time data encoding quantum Boltzmann algorithm where S.P. stands for state preparation, $S_i$ and $C_i$ are the $i$-th streaming and collision operations respectively and M stands for measurement.}
    \label{fig:full_algo}
\end{figure}

Also in general (i.e. $N_t>1$), the streaming step can be implemented by a combination of swap gates. Following the same in-the-vicinity-of-$x$ argument as was used for the collision step, a total of 
\begin{equation}
n_\text{swap}(t) = 4+\sum_{i=1}^{N_t-t }16i = 8 \left ( N_t - t \right )^2 + 8\left ( N_t - t \right ) +4
\end{equation}
swap gates are required to update as many velocity-encoding qubits in time step $t$, whereby these swap operations can be performed largely in parallel.\footnote{In each time step the swap operations in the 4 (or generally speaking $m$) different directions can be performed in parallel. Furthermore the swap operations for the velocities in the same direction but not in the same `line of streaming' can all be performed in parallel. Therefore we only need to take into account the velocities in the same line of streaming and the depth of the circuit is determined by the longest `line of streaming', which is equal to $T-t$. In each layer of the swap operations at least half of the $T-t$ velocities can be swapped to the correct position. Therefore a total of $\log_2 \left (T-t \right )$ swap operations needs to be performed in the $t$-th time step.}
The depth of the streaming circuit at time $t$ will amount to 
\begin{equation}
    d_\text{str}(t) = \log_2 \left ( T-t \right )
\end{equation}
swap operations at time $t$. 
When combining the state preparation with the streaming and collision operations as described the total algorithm can be expressed as in Figure \ref{fig:full_algo}.

\section{Conclusion}
In this paper we have shown that current data encoding methods considered for quantum Boltzmann methods do not allow for treating both streaming and collision as unitary quantum operations. We have provided both a mathematical proof of its impossibility, and insight into the physical properties of the system and encodings that lead to this behavior. Using this insight we subsequently developed a new space-time data encoding method that does allow for both streaming and collision to be implemented as a unitary operation. This paper should serve as a guideline on where (not) to look for successful quantum encodings of the lattice Boltzmann and other QCFD methods.

\printbibliography
\end{document}